\documentstyle[vsolj01,graphicx]{article}

\begin{document}

\title{Detection of an Eclipse during the 1987 Outburst of U Sco}

\author{Taichi Kato}
\author{(Dept. of Astronomy, Kyoto University, Kyoto 606-8502, Japan)}
\email{E-mail: tkato@kusastro.kyoto-u.ac.jp}

\vspace{6mm}

   U Sco is a recurrent nova which showed recorded outbursts in 1863,
1906, 1936, 1979, 1987, and most recently in 1999 (for a summary of the
1987 outburst, see Sekiguchi et al. 1988).  The eclipsing nature of
the object was discovered (Shaefer 1990) after the 1987
outburst.  Shaefer and Ringwald (1995) further obtained eclipse timings
yielding a precise ephemeris.  We examined our time-series photographic
observations obtained during on 1987 May 20, which fortuitously covered
the primary minimum.  This makes the first epoch observation of the
U Sco eclipse during the 1987--1999 outburst interval.

   The observations were made using a 13-cm reflector (fl=1.0m) and Sakura
SR-400 films.  The exposure time was typically 475 sec.  The negatives
and prints were measured by eye, using up to five comparison stars.
The estimated step values were analyzed by the least-squares method
(Kato 1997).  Three comparison stars with known visual magnitudes (AAVSO)
were used to calibrate the zero point and the scale.  The resultant values
are listed in table 1.  There was no considerable systematic difference
between photographically yielded magnitudes and other visual observations
obtained quasi-simultaneously.

\begin{table}[b]
\begin{center}
Table 1. Photographic observations of U Sco \\
\vspace{6pt}
\begin{tabular}{c c c c c c}
\hline
HJD-2400000 & mag & HJD-2400000 & mag & HJD-2400000 & mag \\
\hline
46936.086 & 12.39 & 46936.138 & 12.50 & 46936.190 & 13.05 \\
46936.089 & 12.52 & 46936.144 & 12.92 & 46936.195 & 12.92 \\
46936.093 & 12.78 & 46936.150 & 12.11 & 46936.201 & 12.92 \\
46936.097 & 13.13 & 46936.155 & 12.31 & 46936.206 & 13.45 \\
46936.102 & 13.06 & 46936.168 & 12.59 & 46936.212 & 13.07 \\
46936.109 & 12.43 & 46936.173 & 12.85 & 46936.218 & 12.53 \\
46936.128 & 12.53 & 46936.179 & 12.84 & & \\
46936.134 & 13.40 & 46936.184 & 12.91 & & \\
\hline
\end{tabular}
\vskip 6pt
\end{center}
\end{table}

   Figure 1 shows the overall light curve of U Sco during the 1987 outburst,
drawn from the VSOLJ database and the above estimates (the corresponding data
in the VSOLJ database have been superseded by the updated magnitudes).
Superimposed on the general smooth fading, a temporary fading is apparent
(at HJD 2446936), corresponding to the eclipse minimum.
According to the ephemeris by Shaefer and Ringwald (1995), there is
no other eclipses affecting the overall light curve.
Figure 2 presents the enlarged light curve of the time-series photographic
observations.  Although the scatter is relatively large (especially in the
first half of the observation was slightly affected by slightly poor
tracking and occasionally truncated exposures), there is a clear fading
in the last half of the observation.  The exposure centered on
HJD 2446936.206 shows the object markedly faint, possibly catching
the mid-eclipse.

\begin{figure}
  \begin{center}
  \includegraphics[angle=0,height=9cm]{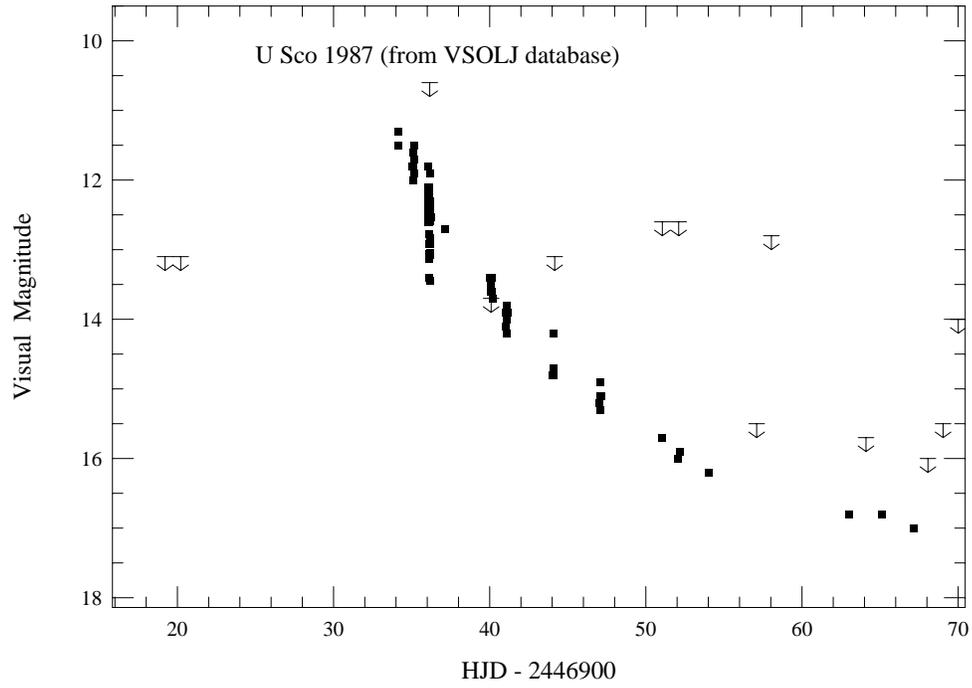}
  \caption{Light curve of the 1987 outburst of U Sco (VSOLJ)} \label{fig-1}
  \end{center}
\end{figure}

   By using the VSOLJ database between 1987 May 18 and 24, and subtracting
the smooth fading trend, all available positive observations (using the same
comparison sequence) were converted to the orbital phases based on
the ephemeris by Shaefer and Ringwald (1995).
The resultant modulations were averaged to 0.02 phase bins (figure 3),
on which the primary eclipse with a depth of $\sim$0.8 mag is clearly
seen.  The relative insufficiency of the points during the egress phase
makes determining the accurate eclipse timing from the phase-averaged
light curve relatively difficult, but the still fading trend on the
time-resolved light curve at the expected minimum (HJD 46936.199) may be
an indication of a slight positive $O-C$.  If we adopt the epoch of the
faintest estimate (HJD 2446936.206), $O-C$ becomes +0.007 d.

\begin{figure}
  \begin{center}
  \includegraphics[angle=0,height=9cm]{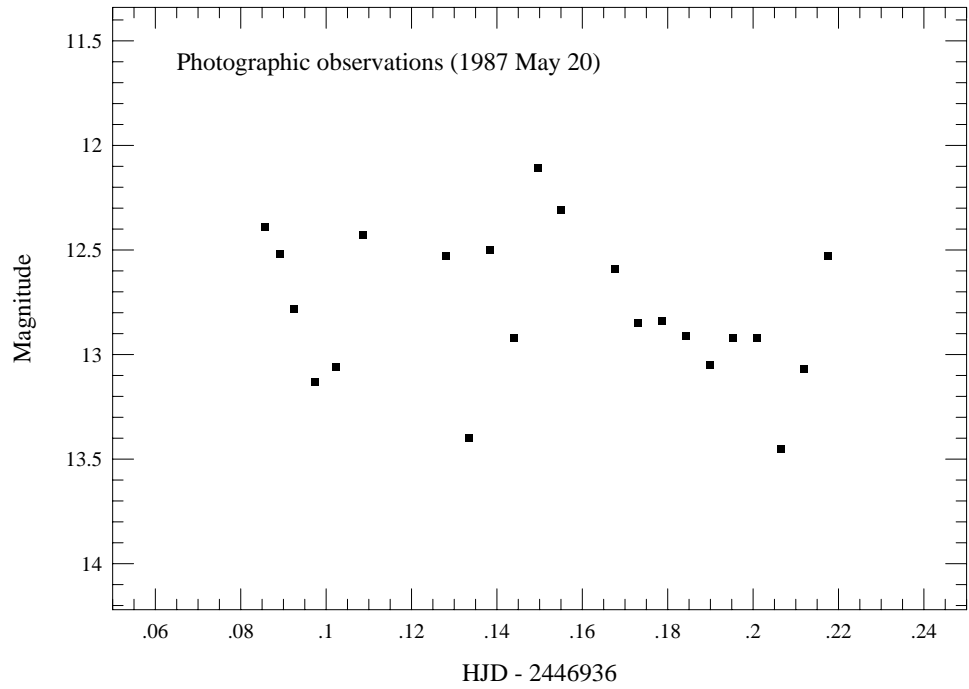}
  \caption{Enlarged light curve on 1987 May 20} \label{fig-2}
  \end{center}
\end{figure}

\begin{figure}
  \begin{center}
  \includegraphics[angle=0,height=9cm]{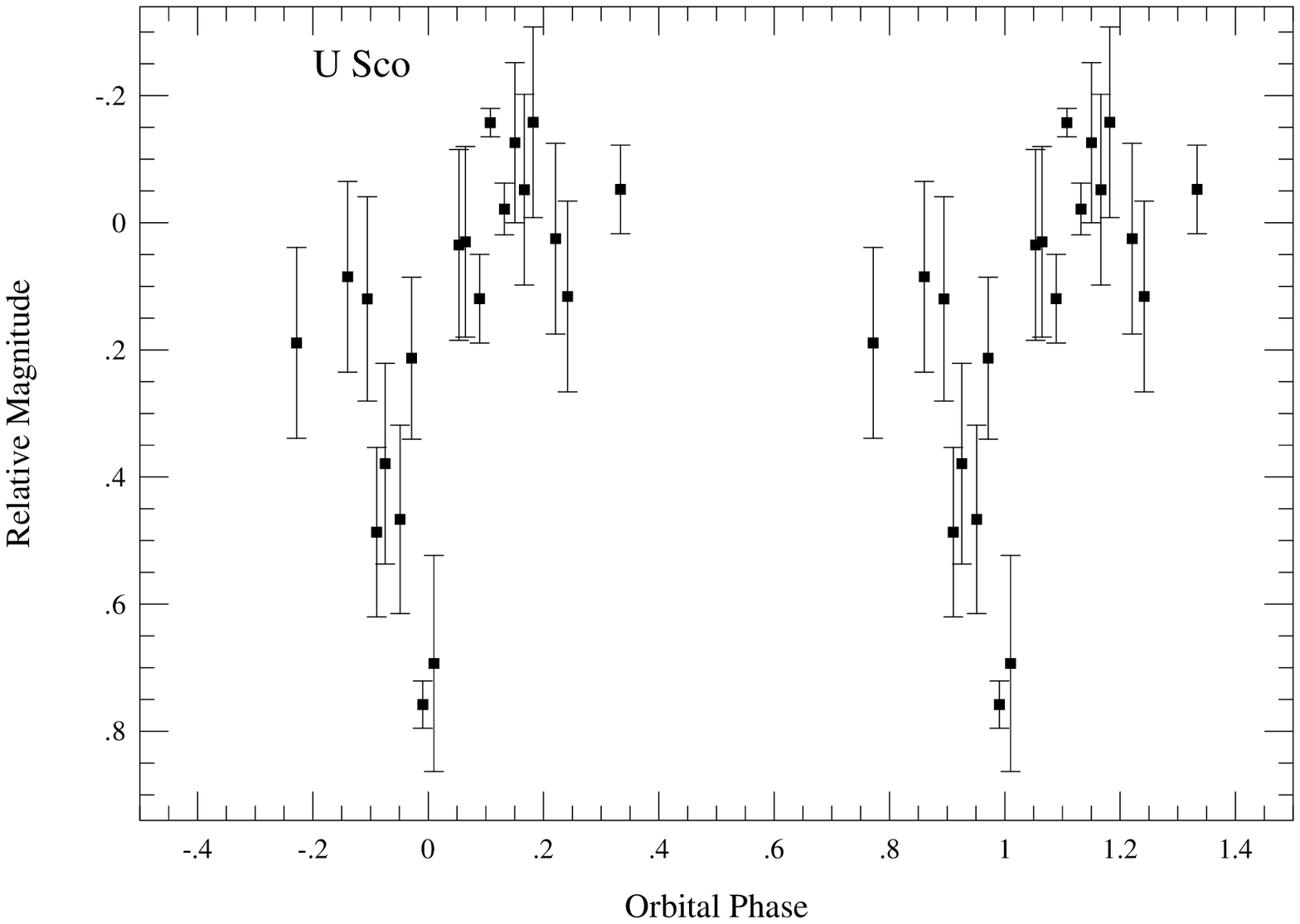}
  \caption{Phase-averaged light curve of the U Sco eclipse} \label{fig-3}
  \end{center}
\end{figure}

   The author is grateful to the members of Kyoto University Astronomy
Lovers' Association, who helped the observations and kindly allowed
the usage of the instruments.  The author is also grateful to the VSOLJ
members and staffs who have made the database available.

\end{document}